\documentclass[aps,reprint,prc,showpacs,preprintnumbers,amsmath,amssymb,nofootinbib]{revtex4-2}
\usepackage{graphicx}
\usepackage{dcolumn}
\usepackage{amsmath}
\usepackage{times}
\usepackage{mathtools}
\usepackage{placeins}
\usepackage{booktabs}
\usepackage{multirow}
\usepackage{makecell}
\usepackage{xcolor}

\usepackage{hyperref}
\usepackage{cleveref}
\usepackage{enumitem}
\usepackage{physics}
\usepackage{leftidx}
\usepackage{subcaption}
\usepackage{tikz}

\usepackage[normalem]{ulem}

\usepackage{multirow} 
\usetikzlibrary{decorations.markings,decorations.pathreplacing,calligraphy}
\usepackage{dsfont}
\usepackage{ragged2e}
\usepackage{orcidlink}

\begin{document}

\title{Coulomb-Induced \boldmath{$K^- pp$} Molecular State}

\author{T. Massimino\, \orcidlink{0009-0004-1303-0759}, Z. Papp \orcidlink{0000-0003-4990-2454}  }
\affiliation{%
Department of Physics, California State University, Long Beach, California 90840}%

\author{J. R\'evai}
\affiliation{%
Budakeszi, Hungary, H-2092}%

\author{J.-M. Richard\,\orcidlink{0000-0001-6459-765X}}
\affiliation{
Universit\'e Lyon 1, Institut de Physique des 2 Infinis de Lyon,\\ 
IN2P3-CNRS, 4, rue Enrico Fermi, Villeurbanne, France}

\author{N.V. Shevchenko\,
\orcidlink{0000-0001-9619-1590}}%
 \email{shevchenko@ujf.cas.cz}
\affiliation{%
 Nuclear Physics Institute of the Czech Academy of Sciences, 25068 \v{R}e\v{z}, Czech Republic}%

\date{\today}

\begin{abstract}

We study a new bound state in the three-body system consisting of an antikaon and two protons. 
The state is primarily induced by Coulomb interactions between the $K^-$ and the two protons, and is modified by strong $K^- p$ and $pp$ interactions. 
This spatially extended molecular state has binding energy and width that are of the order of keV, 
compared to the tens of MeV binding energy and width of the much-studied quasi-bound $K^- pp$ state caused primarily by strong interactions.

\end{abstract}



\maketitle

\section{Introduction}
\label{intro_sect}

The two-body antikaon-proton $K^{-}p$ system is notable as an experimentally studied system whose strong and Coulomb interactions are both attractive and can each separately support (quasi-)bound states.
These interactions are usually treated separately, but in nature, the strong and Coulomb interactions act simultaneously. 
The $\Lambda(1405)$ resonance is usually considered to be a quasi-bound state of an antikaon and nucleon interacting via the strong interaction. 
The Coulomb interaction is usually neglected in $\Lambda(1405)$ calculations.
The pure Coulomb attraction between the negatively charged $K^-$ and the positively charged proton leads to the formation of an exotic atom, kaonic hydrogen.  
Although the strong $\bar{K}N$ interaction plays a secondary role in kaonic hydrogen, this interaction shifts the lowest atomic level and gives it a non-zero width.
These shifts have been measured and are used to study the strong antikaon-nucleon interaction. 

Due to the attractive nature of the antikaon-nucleon strong interaction,
it was assumed that quasi-bound states could exist in the three-body $\bar{K}NN$ system and other few-body antikaonic-nucleus systems \cite{AY}. 
Many theoretical calculations have been performed, mostly focusing on the lightest $K^- pp$ system. 
The predicted binding energies and widths of the quasi-bound state varied substantially, due to different computational methods and two-body inputs. These theoretical predictions agree only on the existence of the quasi-bound state in the $K^- pp$ system.

The first Faddeev calculations of the  $I = 1/2$, $J^{\pi} =0^-$
quasi-bound state in the $\bar{K}NN - \pi \Sigma N$ system, corresponding to the $K^- pp$ in particle representation, were
performed in Refs.~\cite{NGM,NGMR,IS}. The most recent Faddeev calculations~\cite{FineTune} of the system were performed
with three coupled channels $\bar{K}NN$,  $\pi \Sigma N$, and $\pi \Lambda N$ using energy-independent and energy-dependent phenomenological 
and chirally motivated $\bar{K}N - \pi \Sigma - \pi \Lambda$ potentials, respectively, as an input. 

Several experiments hinted at the possible existence of the strong $K^- pp$ quasi-bound state, and recently the E15 experiment at J-PARC~\cite{E15_JPARC1} reported the first clear signal of
the $\bar{K}NN$ quasi-bound state. 
The theoretical results of Ref.~\cite{FineTune} evaluated with one of the $\bar{K}N$ interaction models are in agreement with the experimental results for both the binding energy and width of the quasi-bound state, while other theoretical predictions tend to underestimate the width of the $K^- pp$ state. 

The Coulomb interaction in the strong-induced $K^- pp$ state is usually assumed to give small corrections compared to the binding energy, and hence it is usually omitted. However, it is known that any \(m^\pm M^\mp M^\mp\) system is bound by Coulomb forces, irrespective of the mass ratio $M/m$~\cite{1977JMP....18.2316H}. Therefore, in analogy with the two-body \(K^-p\) case, the $K^- pp$ system can support a bound state predominantly induced by the Coulomb interaction between the negatively charged \(K^-\) and the two positively charged protons. The strong antikaon–nucleon and nucleon–nucleon interactions are expected to play a secondary role in this ``molecular'' \(K^-pp\) state, inducing a shift of the lowest energy level and generating a finite width, in close analogy to the $K^-p$ system.

The present work is devoted to the analysis of this three-body ``molecular'' state in the $K^-pp$ system. Relatively few studies have addressed three-body ions 
$m^\pm M^\mp M^\mp$ in which the Coulomb interaction is supplemented by short-range potentials. Preliminary investigations of three-body systems with very long range and short range interactions include \cite{Richard:2016eca,Moriya:2019ans}.
Here we study the $K^- pp$ system using Faddeev equations in the Coulomb-Sturmian basis with the strong and Coulomb interactions directly taken into account.

Section \ref{two-body} is devoted to the description of the $K^- p$ two-body subsystem of the three-body system. The 
method to solve the three-body problem is described in Section \ref{3formalism}, and the convergence of the three-body energies is discussed in Section \ref{convergence}. The obtained results are presented and discussed in Section \ref{results}, followed by conclusions in Section \ref{conclusion}.

\section{The \boldmath{$K^- p$} Two-Body (Sub)system}
\label{two-body}

Antikaon-nucleon interaction models are usually fitted to the $1s$ level shift and width of (anti)kaonic hydrogen, $K^- p$ cross-sections and threshold branching ratios. 
The data on the $K^- p$ cross-sections are quite old and inaccurate.
The threshold branching ratios are more accurate, but there are only three of them. 
The most recent data on antikaonic hydrogen were measured by the \mbox{SIDDHARTA} collaboration \cite{SIDDHARTA}. 
In Ref.~\cite{FineTune},
different $V_{\bar{K}N}$ potentials were constructed with $\bar{K}N - \pi \Sigma - \pi \Lambda$ coupled channels that reproduce these experimental data equally well. 
They are phenomenological potentials with one- or two-pole structure of the $\Lambda(1405)$ resonance, and a chirally-motivated potential with two-pole structure. 
In this work, we model the $S$-wave strong $K^- p$  
interaction with a complex one-term separable potential
\begin{equation}
	v^{K^- p}  =   \ket{f} \lambda \bra{f},
    \label{eq:VKp}
\end{equation}
with Yamaguchi form factors \cite{Yamaguchi1954}, defined in momentum space by
\begin{equation}
	\braket{\textbf{k}}{f} = (k^2 + \beta^2)^{-1},
    \label{eq:Vff_k}
\end{equation}
and configuration space by
\begin{equation}
    \braket{\textbf{r}}{f} = \sqrt{\frac{\pi}{2}} \frac{e^{-\beta r}}{r}.
    \label{eq:Vff_r}
\end{equation}
This potential with strength constant $\lambda=-2.460-i\,0.537~\mbox{fm}^{-2}$
reproduces the binding energy of $\Lambda(1405)$ from the one-pole three-channel phenomenological potential $V_{\bar{K}N-\pi \Sigma - \pi \Lambda}^{\rm 1,SIDD-A}$ from Ref.~\cite{FineTune} with an identical range parameter 
$\beta=3.95\ \mbox{fm}^{-1}$. 

The two-body bound-state energies of the $K^- p$ system are the solutions of the homogeneous Lippmann-Schwinger equation,
\begin{equation}
    |\psi\rangle = g_l^{C}(\varepsilon) v^{K^- p} |\psi \rangle,
    \label{LS0}
\end{equation}
where $g_l^{C}(\varepsilon)$ is the Coulomb Green's operator and $l$ is the angular momentum. Eq.~(\ref{LS0}) with the separable potential (Eq.~\ref{eq:VKp}) and form-factors (Eq.~\ref{eq:Vff_k}) can be solved exactly. Since we performed our three-body calculations using the Coulomb-Sturmian (CS) representation \cite{PhysRevC.38.2457, PhysRevA.46.4437,Konya:1999by,papp2026mastering}, we solved the two-body equation for the $K^- p$ system using the CS representation as well and compared the two-body results with the exact ones.

The CS basis in configuration space is defined by
\begin{equation}
\begin{split}
    \langle r|nl;b\rangle &= \left( \frac{n!}{(n+2l+1)!} \right)^{1/2} e^{-br}(2br)^{l+1}L_n^{(2l+1)}(2br) \\
    \langle r|\widetilde{nl;b}\rangle &= \langle r|nl;b\rangle /r,
\end{split}
\end{equation}
where $L_n^{(\alpha)}$ are the associated Laguerre polynomials,
and $b$ is a parameter.
The CS states satisfy biorthogonality and completeness relations, 
\begin{equation}
    \langle nl;b|\widetilde{n'l;b}\rangle = \delta_{n n'},
    \qquad 
    \sum_{n=0}^{\infty}|\widetilde{nl;b}\rangle \langle nl;b| = {1},
\end{equation}
and form a basis in the $l$  angular momentum channel.

In this solution method, we approximate the potential as
\begin{equation}
    v^{K^- p} \approx \sum^N_{n,n'=0} \ket*{\widetilde{nl;b}} \underline{v}^{K^-p} \bra*{\widetilde{n'l;b}},
\end{equation}
where $\underline{v}^{K^-p} = \mel{nl;b}{ v^{K^-p}}{n'l;b}$. The CS matrix elements of the potential with
$l=0$ Yamaguchi form factors can be given in closed form
\begin{equation}
  \underline{v}^{K^-p}  = \lambda \braket{n0;b}{f} \braket{f}{n'0;b}  
\end{equation}
with 
\begin{equation}
    \braket{n0;b}{f} = 2\pi \sqrt{2(n+1)} b(\beta-b)^n/(\beta +b)^{n+2}.
\end{equation}

This approximation of the potential allows the Lippmann-Schwinger equation~\eqref{LS0} to be written as
\begin{equation}
    |\psi\rangle = \sum^N_{n,n'=0}  g_l^{C}(\varepsilon) \ket*{\widetilde{nl;b}}   \underline{v}^{K^-p}
    \braket*{\widetilde{n'l;b}}{\psi },
    \label{LS1}
\end{equation}
where the summation is truncated at finite $N$.
Acting with a truncated set of bra states $\bra*{\widetilde{m l;b}}$, 
the Lippmann-Schwinger equation becomes a matrix equation,
\begin{equation}
    \underline{\psi} = \underline{g}_l^{C}(\varepsilon) \underline{v}^{K^-p} \underline{\psi},
\end{equation}
which is in fact a homogeneous linear algebraic equation where the solutions are given by the condition
\begin{equation}
    \det( ( \underline{g}_l^{C}(\varepsilon))^{-1} - \underline{v}^{K^-p} )   = 0,
\end{equation}
with $\underline{g}_l^{C}(\varepsilon)= \mel*{\widetilde{nl;b}}{ g_l^{C}(\varepsilon)} {\widetilde{n'l;b}}$ 
and $\underline{\psi} = \braket*{\widetilde{nl;b}}{\psi}$.

In the CS basis the matrix elements of the short-range potential can easily be calculated, at least numerically, but 
for Yamaguchi form factors they are given in closed form. The  matrix elements of Coulomb Green's operator can be given 
in terms of $\mbox{}_2F_1$ hypergeometric functions \cite{PRADemir2006, papp2026mastering}.

The results of our studies of the two-body $K^-p$ system were reported in  \cite{kpPreprint} for real and complex strong potentials.
Table \ref{tbl:2b_results} shows
the energies of the Coulomb-only and complex strong-only states alongside the two lowest states of the complex strong plus Coulomb potentials.
The strong-induced quasi-bound state ($\varepsilon_{1s}^{s+C}$) corresponds to the $\Lambda(1405)$ resonance, and the Coulomb-induced atomic state ($\varepsilon_{2s}^{s+C}$) corresponds to the lowest level of kaonic hydrogen. 
The wave functions of these two states, calculated with a real potential, are shown in Fig.~2 of Ref.~\cite{kpPreprint}, 
confirming that the lowest level of kaonic hydrogen is an excited state of the $K^- p$ spectrum, as it has a node near the origin.
The two states $\varepsilon_{1s}^{s+C}$ and $\varepsilon_{2s}^{s+C}$ have vastly different length scales. 

\begin{table}[ht]
\caption{\justifying 
Two-body energies of the $K^- p$ system calculated with Coulomb-only, strong-only, and strong plus Coulomb potentials \cite{kpPreprint}.
}
\label{tbl:2b_results}
\begin{ruledtabular}
\begin{tabular}{clcc}
\textbf{System} & \textbf{Level} & \multicolumn{2}{c}{\textbf{Two-Body Energy}} \\[1mm]
\noalign{\smallskip} \hline
strong-only & $\varepsilon_0^{s}$ & $-5.37 - i \, 33.50$ & MeV \\
\noalign{\smallskip} \hline
Coulomb-only & $\varepsilon_{1s}^{C}$ & $-8.64$ & keV \\
\noalign{\smallskip} \hline
\multirow{2}{*}{strong\,+\,Coulomb} &
  $\varepsilon_{1s}^{s+C}$ & $-7.68 -i \, 34.40$ & MeV \\
 & $\varepsilon_{2s}^{s+C}$ & $-8.13 - i \, 0.25$ & keV \\
\end{tabular}\end{ruledtabular}
\end{table}

A level rearrangement effect was observed in the $K^-p$ system.
By gradually varying the strength parameter of the strong interaction, we demonstrated that the $1s$ and $2s$ states of the Coulomb-only system continuously transform into the $\Lambda(1405)$ resonance and the lowest state of kaonic hydrogen, respectively.

\section{The \boldmath{$K^-pp$} three-body system}
\label{3formalism}

For the three-body problem we work in the frame of Jacobi coordinates $x_\alpha, y_\alpha$,
where $\alpha=1,2,3$, and $x_\alpha$ is the coordinate between the two other particles $\beta$
and $\gamma$, and $y_\alpha$ connects the center of mass of the $(\beta,\gamma)$ pair and $\alpha$.
We index the antikaon with $3$ and the two protons with $1$ and $2$, respectively.

In the $K^-pp$ system at low energies
the proton-proton subsystem is an $S$ wave, isospin triplet, and spin singlet state.
We model the strong interaction between protons with
the MT-I Malfliet-Tjon potential \cite{Malfliet1969}
\begin{equation}
    v^{MT-I}(r) = -V_A \frac{e^{-\mu_A r}}{r} + V_R \frac{e^{-\mu_R r}}{r},
\end{equation}
where $V_A = 513.968\ \mbox{MeV fm}$, $V_R = 1438.720\ \mbox{MeV fm}$, 
$\mu_A = 1.55\ \mbox{fm}^{-1}$, 
and
$\mu_R = 3.11\ \mbox{fm}^{-1}$. 
The full $pp$ potential, $v_3$, is given by a repulsive Coulomb potential plus the MT-I nuclear potential,
\begin{equation}
    v_3(x_3) = v^C(x_3) + v^{MT-I}(x_3).
\end{equation}
The full $K^-p$ potentials, $v_1$ and $v_2$, are given by the sum of an attractive Coulomb term and an attractive strong potential,
\begin{equation}
    v_1(x_1) = v^C(x_1) + v^{K^- p}(x_1). 
\end{equation}

While  $v^{K^-p}$ is a genuine short range potential, the attractive  Coulomb potential has both long range and short range
features. On the one hand, it is a short range potential as it supports bound states.
On the other hand, it is a long range potential as it alters the asymptotic motion and supports infinitely many bound states. 
We split the potential into short- and long-range parts,
$v_\alpha = v_\alpha^{(s)} + v_\alpha^{(l)}$.
The choice of splitting can be arbitrary,
as long as 
the long-range part contains the proper asymptotics
of the two-body system without supporting bound states.
This ensures the absence of spurious solutions in the Faddeev equation.
We adopt Merkuriev's splitting \cite{merkuriev1980three,faddeev2013quantum}, 
\begin{equation}
    \begin{split}
    v_1^{(s)}(x_1,y_1) & = v^C_1(x_1) \cdot \zeta(x_1,y_1) + 
 v^{K^- p}_1(x_1), \\
    v_1^{(l)}(x_1,y_1) & = v^C_1(x_1,y_1) \cdot [1-\zeta(x_1,y_1)],
    \end{split}
\end{equation}
where $\zeta(x,y) = 2/\big[1+\mathrm{exp}\big((x/x_0)^\nu/(1+y/y_0)\big)\big]$ 
separates the potential in three-body configuration space. Here $\nu > 2$ and $x_0$ and $y_0$ are chosen to match the size of the system. 
The long range potential $v^{(l)}$ is a valley that becomes wider and shallower as $y\to \infty$ ensuring no bound state
in that limit and consequently no two body asymptotic channel. 

Since the $pp$ system does not have any bound states, 
the entire $v_3$ potential can be treated as long range type.
Thus, there is no associated asymptotic channel, 
and there is no need to devote a Faddeev component to this configuration  \cite{papp2002resonant,papp2002faddeev}.

The three-body Hamiltonian  is given by
\begin{equation}
    H = H^{(l)} + v^{(s)}_1 + v^{(s)}_2,
\end{equation}
where the long range Hamiltonian is
\begin{equation}
    H^{(l)} = H^{(0)} + v^{(l)}_1 + v^{(l)}_2 + v_3.
\end{equation}
Here,
$H^{(0)}$ is the kinetic energy operator,
and
$v^{(l)}_\alpha(x_\alpha)$ and
$v^{(s)}_\alpha(x_\alpha)$  
with $\alpha = 1,2$, are the interparticle long- and short-range potentials, respectively.

Hence, the Faddeev decomposition reads
\begin{equation}
    |\Psi\rangle = |\psi_1\rangle + |\psi_2\rangle,
\end{equation}
where
\begin{equation}
    |\psi_1\rangle= G^{(l)} (E) v_1^{(s)}|\Psi\rangle
         \mathrm{~~and~~}
    |\psi_2\rangle= G^{(l)}(E)v_2^{(s)}|\Psi\rangle,
\end{equation}
with $G^{(l)}(z) = (z-H^{(l)})^{-1}$.

The Faddeev decomposition of the wave function results in a set of coupled Faddeev integral equations
\begin{align}
    \begin{split}
    |\psi_1\rangle & = G^{(l)}_1 (E) v^{(s)}_1 |\psi_2\rangle \\
    |\psi_2\rangle &= G^{(l)}_2 (E) v^{(s)}_2 |\psi_1\rangle,
    \end{split}
\end{align}
where $G^{(l)}_\alpha(z) = (z-H^{(l)} -v^{(s)}_\alpha)^{-1}$.

Particles $1$ and $2$ are identical. Consequently, the Faddeev components $\ket{\psi_1}$ and $\ket{\psi_2}$, in their own natural
Jacobi system, have the same functional form.
So, $\ket{\psi_2}$ can be expressed from $\ket{\psi_1}$ by interchanging the identical 
particles, $|\psi_2\rangle = p{\cal P}|\psi_1\rangle$.
Here, $\cal P$ is the permutation operator and $p = \pm 1$ is the eigenvalue of $\cal P$.
This simplifies the Faddeev equations to a single equation,
\begin{equation}
    \ket{\psi_1 } = G^{(l)}_1 (E) v^{(s)}_1 p {\cal P} \ket{\psi_1}.   
    \label{fadd1}
\end{equation}

We solve the three-body problem by using the Faddeev method in Coulomb-Sturmian basis representation \cite{PhysRevC.54.50,PhysRevC.55.1080,PhysRevA.63.062721}. 
For three-body problems the appropriate CS basis is the direct  product of two-body Hilbert space bases
\begin{equation}
    \langle xy |n\nu  \rangle_\alpha = 
    \langle x|nl\rangle_\alpha \otimes \langle y|\nu\lambda \rangle_\alpha, 
\end{equation}
where the angular momentum coupling is implicitly assumed.
The solution of Eq.\ \eqref{fadd1} follows the same logic as in the two-body case.
We approximate the short range potential in the three-body basis by
\begin{equation}
    \ket{\psi_1 } = 
    \sum^N_{n \nu n' \nu'}
    G^{(l)}_1 
    |\widetilde{n\nu}\rangle_1  
    \mbox{}_1 \langle n\nu|
    v_1^{(s)}p{\cal P}
    |n'\nu' \rangle_1  
    \mbox{}_1\langle \widetilde{n'\nu'}
    |\psi_1\rangle .   
    \label{fadd1v}
\end{equation}

As we did previously, we act with $\mbox{}_1\langle \widetilde{n''\nu''}| $ from the left and obtain a matrix
equation, 
\begin{equation}
    \left(\left[\underline{G}_1^{(l)} (E)\right]^{-1} 
    - \underline{v}^{(s)}_1\right) 
    \underline{\psi}_1
    =0,
\end{equation}
where 
\begin{align}
    \underline{v}^{(s)}_1 &= 
    \mbox{}_1 \langle n\nu  |
    v_1^{(s)}p{\cal P}
    |n'\nu'  \rangle_1  = p \cdot \mbox{}_1 \langle n\nu  |
    v_1^{(s)}
    |n'\nu' \rangle_2
    \\
    \underline{G}^{(l)}_1 &= 
    \mbox{}_1 \langle \widetilde{n\nu}|
    {G}^{(l)}_1
    |\widetilde{n'\nu'} \rangle_1
    \\
    \underline{\psi}_1 &=
    \mbox{}_1 \langle\widetilde{ n\nu }
    |\psi_1\rangle.
\end{align}

We can calculate $\mbox{}_1 \langle n\nu  | v_1^{(s)} |n'\nu' \rangle_2$ numerically by transforming the CS states between different sets of 
Jacobi coordinates \cite{balian1969angular,papp2026mastering}. 
However, the calculation of 
$\underline{G}^{(l)}_1 =  \mbox{}_1 \langle \widetilde{n\nu}  | {G}^{(l)}_1 |\widetilde{n'\nu'} \rangle_1$ poses some challenges, 
as it is the resolvent of a rather complicated three-body Hamiltonian
\begin{equation}
    G_1^{(l)}(z) = (z - H_1^{(l)} )^{-1},
\end{equation}
where 
\begin{equation}
    H_1^{(l)}  = H^{(0)} + v_1 + v_2^{(l)}+ v_3.
\end{equation}
Here $H^{(0)}= h^{(0)}_{x_1} + h^{(0)}_{y_1}$ is the three-body kinetic energy, which can be expressed in terms of two-body kinetic energies in any pair of Jacobi coordinates.
To calculate $G_1^{(l)}(z)$, we introduce another three-body Hamiltonian,
\begin{equation}
    \tilde{H}_1(x_1,y_1) = h_{x_1}   + h^{(0)}_{y_1}.
\end{equation}
where $h_{x_1} = h^{(0)}_{x_1} + v_1(x_1)$.
Consequently,  
\begin{equation}
    H_1^{(l)} =  \tilde{H}_1(x_1,y_1) + v_2^{(l)}(x_2,y_2) + v_3(x_3).
    \label{eqtilde}
\end{equation}
From this relation follows the resolvent equation, 
\begin{equation}
    G_1^{(l)}(z) = \tilde{G}_1(z) + \tilde{G}_1(z) (v_2^{(l)} + v_3) G_1^{(l)}(z).
\end{equation}

To solve the Faddeev equation, we need only the CS matrix elements between the basis states $\ket{n \nu}_1$. So, finally we have
\begin{equation}
    \underline{G}_1^{(l)}(z) = ( (\underline{\tilde{G}}_1(z))^{-1} - \underline{U} )^{-1},
\end{equation}
where $\tilde{\underline{G}}_1 = 
    \mbox{}_1 \langle \widetilde{n\nu}  |
    \tilde{G}_1
    |\widetilde{n'\nu'} \rangle_1$
and $\underline{U} = \mbox{}_1 \langle n\nu  | (v_2^{(l)} + v_3)|n'\nu' \rangle_1 $, the latter of which we can evaluate numerically.

\begin{figure}
    \begin{tikzpicture}
        [line width=1pt,
        decoration={%
          markings,
          mark=at position 0.17 with {\arrow[line width=1.2pt]{>}}, 
          mark=at position 0.5 with {\arrow[line width=1.2pt]{>}},
          mark=at position 0.83 with {\arrow[line width=1.2pt]{>}},   
        }
        ]
        \draw [help lines,dashed,->] (-5,0) -- (3,0) coordinate (xaxis);
        \draw [help lines,dashed,->] (0,-3)coordinate(-yaxis) -- (0,2.5) coordinate (yaxis);
        \node [below] at (xaxis) {$\mathrm{Re}(z')$};
        \node [above] at (yaxis) {$\mathrm{Im}(z')$};

        \draw [postaction=decorate, line width=1pt] (-1.25,2.5) -- (-1.25,-3) node[below right, yshift=15pt] {$C$};
        \fill[blue] (-4, -1.1) circle (2pt);
        \draw [postaction=decorate] (-3.8,-1.1) arc (0:360:0.2);
        \fill[blue] (-2.25, -0.1) circle (2pt);
        \draw [dashed,postaction=decorate] (-2.05,-0.1) arc (0:360:0.2);
        \fill[blue] (-1.6, -0.05) circle (2pt);
        \draw [postaction=decorate] (-1.4,-0.05) arc (0:360:0.2);
        \fill[blue] (-1, 0) circle (2pt);
        \fill[blue] (-0.8, 0) circle (2pt);
        \fill[blue] (-0.7, 0) circle (2pt);
        \fill[blue] (-0.6, 0) circle (2pt);
        \fill[blue] (-0.5, 0) circle (2pt);
        \fill[blue] (-0.4, 0) circle (2pt);
        \fill[blue] (-0.3, 0) circle (2pt);
        \fill[blue] (-0.2, 0) circle (2pt);
        \fill[blue] (-0.1, 0) circle (2pt);
        \draw [blue, line width=2pt] (0,0) -- (3,0) 
        node[midway, above] {$g_x(z')$};
        
        \draw [red, line width=2pt] (-5,-0.5) -- (-2,-0.5) 
        node[pos=0.3, above] {$g_y(E-z')$};
        
        \draw [gray,decorate,decoration={brace,amplitude=5pt,mirror}]
        (-2,-0.5) -- (0,-0.5) node[midway,xshift=1.0em,yshift=-1.5em]{$\mathrm{Re}(E)$};
        \draw [gray,decorate,decoration={brace,amplitude=5pt,mirror}]
        (0,-0.5) -- (0,0) node[midway,xshift=2.5em]{$\mathrm{Im}(E)$};
    \end{tikzpicture}
    \caption{\justifying
    Schematic of the contour $C$ (black) used in the convolution integral, Eq.\ (\ref{eq:contour}), corresponding to  
    the three-body molecular state. The bound state spectrum of $g_x$ is 
    deformed due to the complex absorbing potential $v^{K^- p}$.
    The contour encircles the spectrum of $g_x$  (blue) without penetrating the spectrum of $g_y$ (red). 
    The deepest bound state pole of $g_x$ stays on the first Riemann sheet of $g_y$, while the first excited state pole is
    submerged into the second Riemann sheet of $g_y$ (dashed contour).
    }
    \label{fig:contour}
\end{figure}
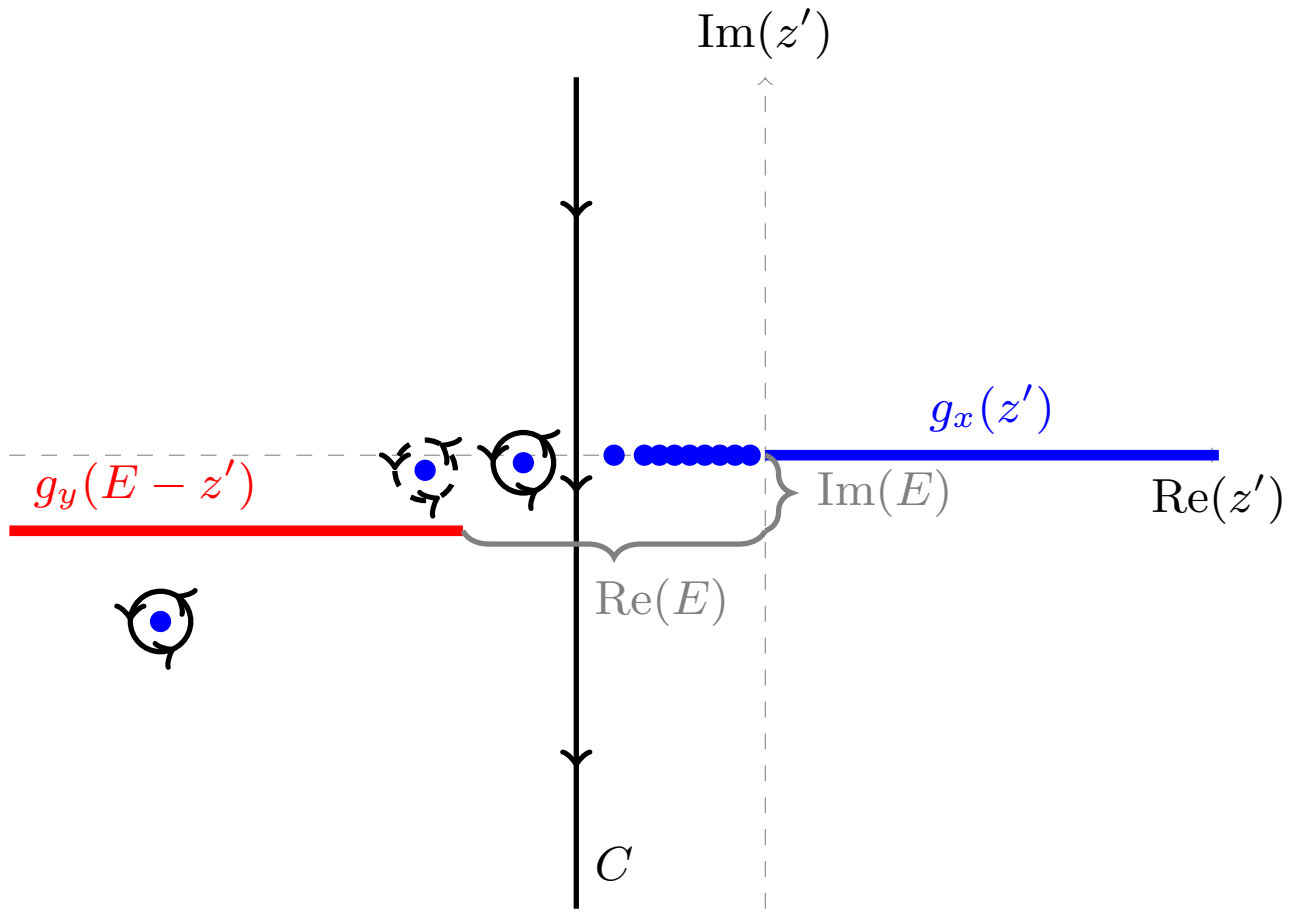

The three-body Hamiltonian $\tilde{H}_1$ is a sum of two commuting two-body Hamiltonians acting in different Hilbert spaces. 
Therefore, its resolvent operator can be given in terms of a convolution integral of two-body Green's operators \cite{bianchi1964convolution,papp2026mastering},
\begin{align}
    \tilde{G}_1(E) &= \frac{1}{2 \pi i} \oint_C dz' g_y(E-z')g_x(z'),
    \label{eq:contour}
\end{align}
where the two-body Green's operators are $g_x(z) = (z - h_x)^{-1}$
and $g_y(z) = (z - h_y^0)^{-1}$.
The contour $C$ is chosen such that it encircles the spectrum of $g_x(z)$ 
without penetrating into the singularities of $g_y(z)$  (Fig.\ \ref{fig:contour}).

The three-body energies and widths of the $K^- pp$ system are given by the zeroes of the Fredholm determinant in the complex plane,
\begin{equation}
    \det\left(\left[\underline{G}_1^{(l)} (E)\right]^{-1} 
    - \underline{v}^{(s)}_1\right)
    =0.
\end{equation}

\section{Convergence of Three-Body Energies in \boldmath{$N$}}
\label{convergence}

In the method described above, 
we approximate the short-range potentials by retaining a finite number of angular momentum channels
and truncating the Coulomb-Sturmian basis at $N$.
We use eight angular momentum channels, 
which can resolve the three-body Coulomb-only system to four digits, see Ref.~\cite{papp2002resonant}.
Given that the $v^{K^-p}$ potential acts only in the $l=0$ angular momentum channel, eight angular momentum channels are more than sufficient.

Usually, one can achieve rapid convergence in $N$
if the CS parameter $b$ is chosen to match the effective size of the states and the effective range of the interactions, 
allowing the truncated basis to resolve the system.
In the $K^- pp$ system, however,
the strong and Coulomb interactions operate on vastly different length scales.
Therefore, the optimal $b$ parameters are vastly different for the strong- and Coulomb-only systems. 
As a result, a rather large basis size is necessary to simultaneously treat both interactions.

\begin{figure}
    \centering
    \begin{subfigure}{0.48\textwidth}
        \centering
        \includegraphics[width=\textwidth]{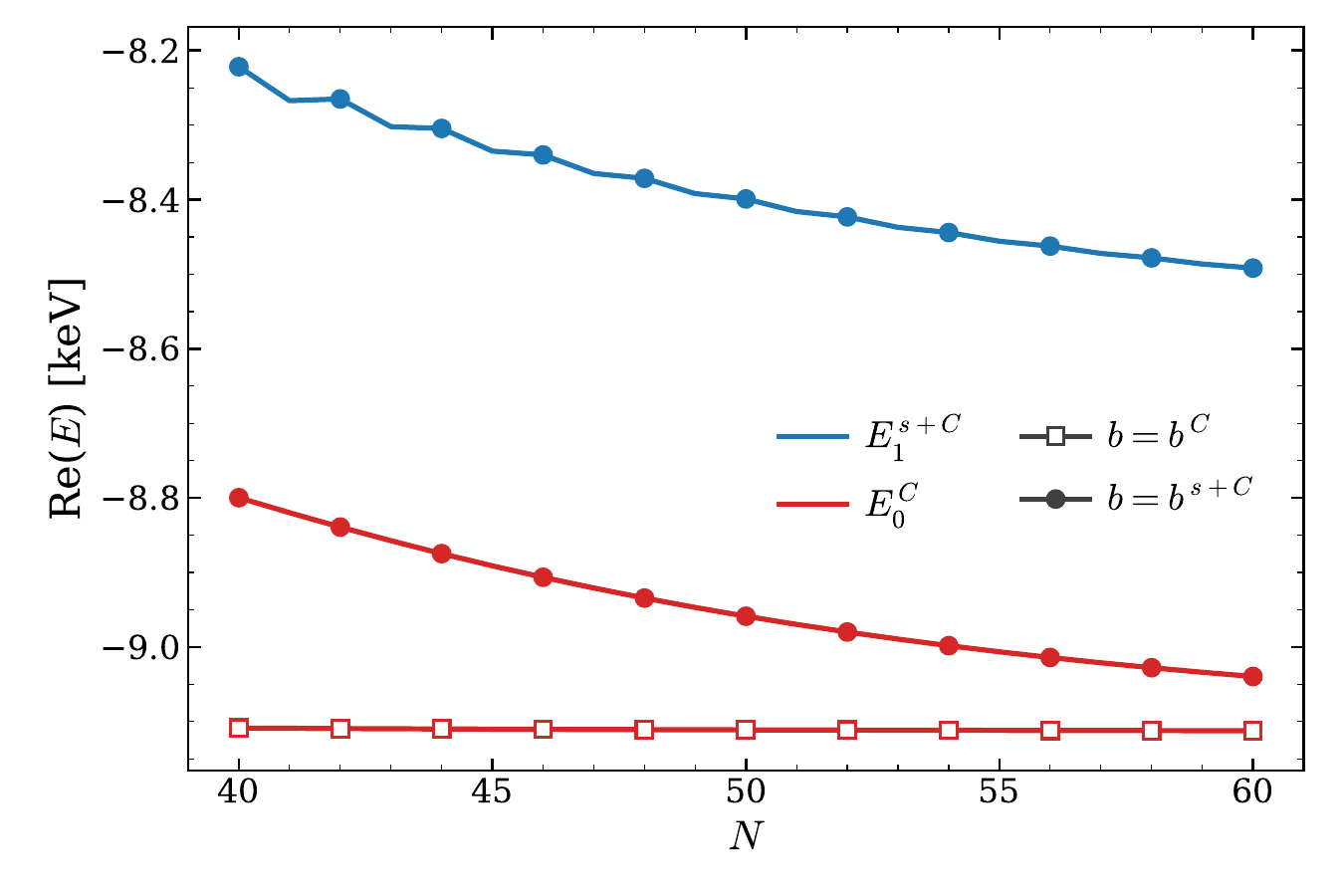}
        \caption{\justifying 
        The ground state energy of the Coulomb-only system (red) and the excited state energy of the strong plus Coulomb system (blue). 
        The parameter $b^{C}$ cannot be used for the latter system, as it does not reproduce the two-body spectrum.
        }
        \label{fig:en3b_convergence_ex}
    \end{subfigure}
    \hfill
    \begin{subfigure}{0.48\textwidth}
        \centering
        \includegraphics[width=\textwidth]{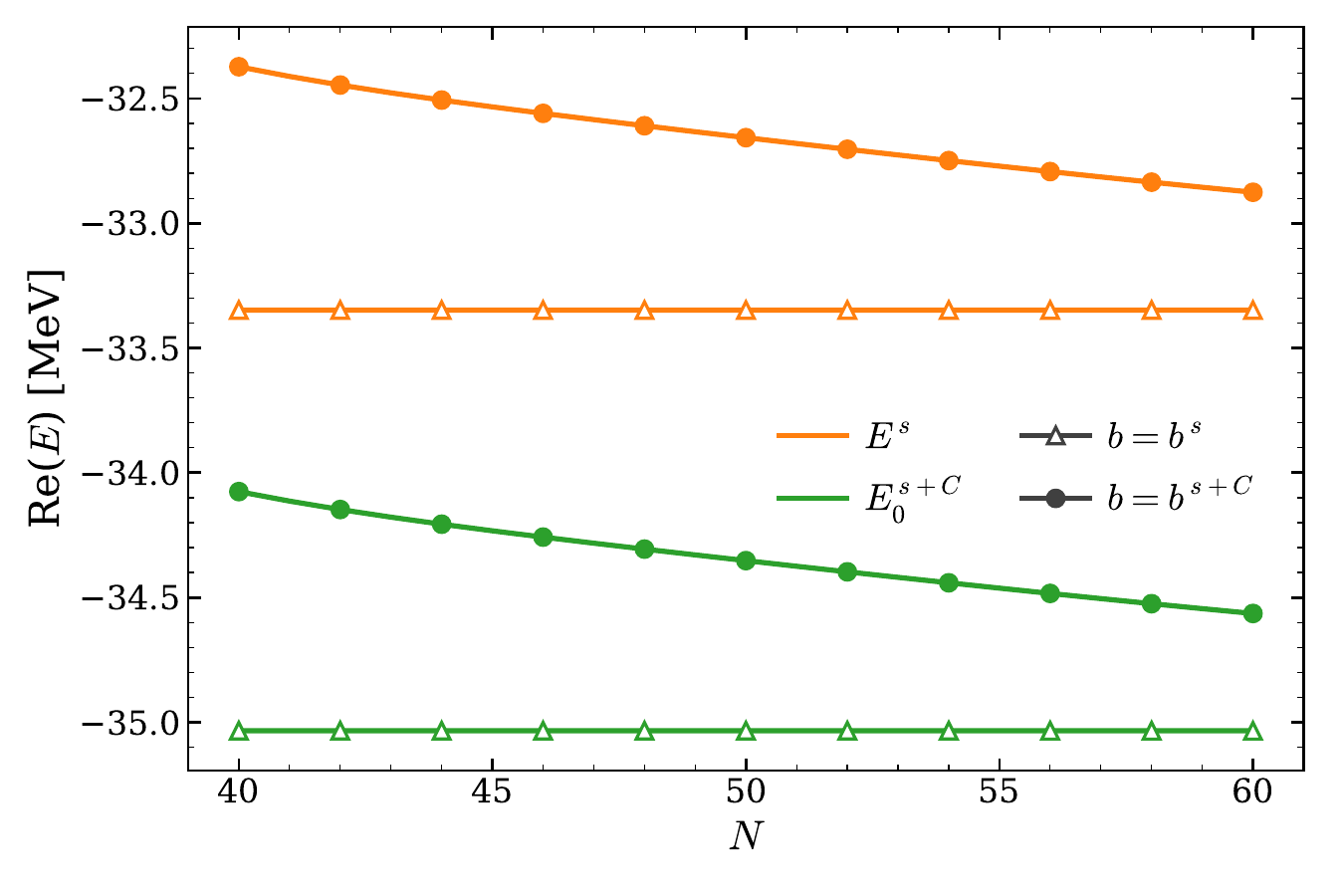}
        \caption{\justifying The ground state energies of the strong-only system (orange) and the strong plus Coulomb system (green). 
        }
        \label{fig:en3b_convergence_gd}
    \end{subfigure}

    \caption{\justifying The convergence of the real part of the three-body energies of $K^- pp$ states.
    Slow convergence occurs for all states when calculated with the intermediate CS parameter, $b^{s+C}$ (filled circles).
    However, $E^s_0$ and $E^{s+C}_0$ converge quickly with $b^s$ (open triangles), and $E^C_0$ converges quickly with $b^C$ (open squares).
    }
    
\end{figure}

As seen in Fig.~\ref{fig:en3b_convergence_ex},
the ground state energy of the Coulomb-only system, $E^C_0$, rapidly converges in $N$ with the choice of $b^{C}= 0.005 \,\mathrm{fm}^{-1}$ (open squares), which creates a spatially extended basis.
However, this basis cannot resolve the short-range strong interaction and gives an incorrect two-body strong plus Coulomb  spectrum $\varepsilon^{s+C}_i$.
As such, it cannot be used to calculate the combined system.
A different choice, $b^s=6.0 \,\mathrm{fm}^{-1}$ (open triangles), creates a spatially compact basis and leads to rapid convergence of the ground state energies of the strong-only system, $E^s_0$, and strong plus Coulomb system, $E^{s+C}_0$ (Fig.~\ref{fig:en3b_convergence_gd}). 
However, it cannot resolve the spatially extended strong plus Coulomb excited state, $E^{s+C}_1$.
The minimum value for the $b$ parameter that reproduces the two-body spectrum correctly is $b^{s+C}=0.2  \,\mathrm{fm}^{-1}$ (filled circles).
This intermediate parameter resolves both interactions, albeit slowly, and can be used to calculate the energies of the combined system, $E^{s+C}_i$.

\begin{figure}
    \centering
    \includegraphics[width=\columnwidth]{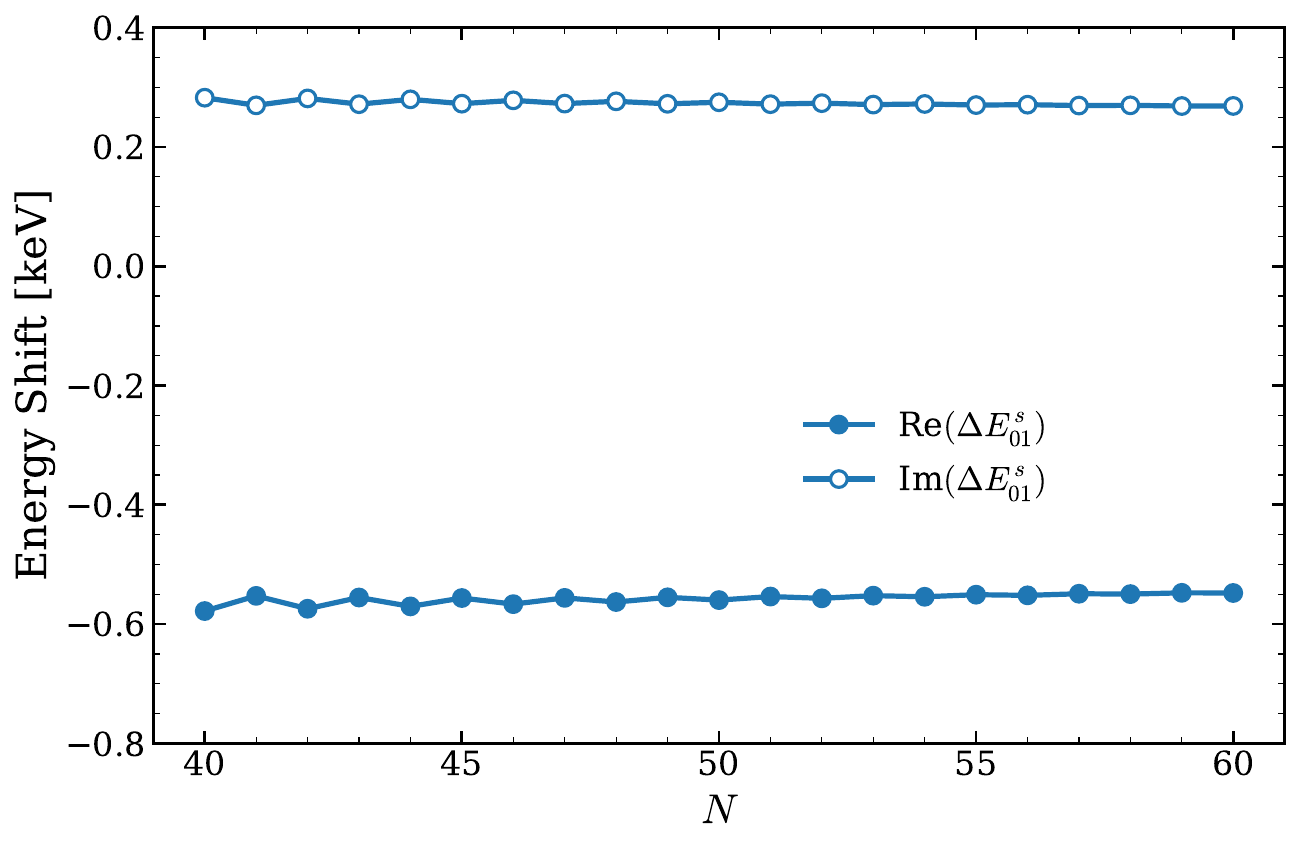}
    \caption{\justifying
    The stability of the real (filled circles) and imaginary (empty circles) parts of the energy difference $\Delta E_{01}^s$ 
    as a function of $N$,
    calculated with $b=b^{s+C}$.
    The energy differences are more stable
    than the energies themselves (Figs.~\ref{fig:en3b_convergence_ex},\ref{fig:en3b_convergence_gd}).
    }
    \label{fig:shifts}
\end{figure}

No choice of $b$ leads to quick convergence of $E^{s+C}_1$.
While
the energies are slow to converge with $b^{s+C}=0.2\,\mathrm{fm}^{-1}$, we note that the energy difference
between the energies of the Coulomb-only ground state and the Coulomb-induced molecular state,
\begin{equation}
\Delta E_{01}^s = E_0^C - E_1^{s+C}
\label{eq:Deltas}
\end{equation}
is more stable than the energy levels themselves, see Fig.~\ref{fig:shifts}.
Thus, we can obtain the excited state energy of the strong plus Coulomb system as a difference between two converged quantities by shifting the energy of the Coulomb-only state calculated with $b^{C}$ by the energy difference calculated with $b^{s+C}$,
\begin{eqnarray}
    E^{s+C}_1 &\approx& E^C_0|_{b=b^C} - \Delta E_{01}^s|_{b=b^{s+C}}.
    \label{eq:approx1}
\end{eqnarray}

We can apply a similar extrapolation to obtain the ground state energy,
\begin{eqnarray}
    E^{s+C}_0 &\approx&  E^s_0|_{b=b^s} - \Delta E_0^C|_{b=b^{s+C}},
    \label{eq:approx0}
\end{eqnarray}
where the energy difference between the strong-induced quasi-bound state without and with the added Coulomb interaction is given by
\begin{equation}
\Delta E_0^C = E_0^s - E^{s+C}_0.
\label{eq:DeltaC}
\end{equation}

\section{Results \& Discussion}
\label{results}

The energies and widths of the three-body $K^-pp$ system with Coulomb-only ($E_0^{C}$), strong-only ($E_0^{s}$), and combined strong plus Coulomb interactions ($E_0^{s+C}$ and $E_1^{s+C}$) are presented in Table \ref{tbl:3b_results}. 
The three-body energies $E_0^{s+C}$ and $E_1^{s+C}$ were obtained using Eqs. (\ref{eq:approx0}) and (\ref{eq:approx1}), respectively.

\begin{table}[ht]
\caption{\justifying 
Three-body energies of the $K^- pp$ system calculated with Coulomb-only, strong-only, and strong plus Coulomb potentials.
}
\label{tbl:3b_results}
\begin{ruledtabular}
\begin{tabular}{clcc}
\textbf{System} & \textbf{Level} & \multicolumn{2}{c}{\textbf{Three-Body Energy}} \\[1mm]
\noalign{\smallskip} \hline
strong-only & $E_0^{s}$ & $-33.35 -i \, 52.73$ & MeV \\
\noalign{\smallskip} \hline
Coulomb-only & $E_{0}^{C}$ & $-9.11$ & keV \\
\noalign{\smallskip} \hline
\multirow{2}{*}{strong\,+\,Coulomb} & $E_0^{s+C}$ & $-35.04 -i \,53.18$ & MeV \\
 & $E_1^{s+C}$ & $-8.56 -i \, 0.27$ & keV \\
\end{tabular}
\end{ruledtabular}
\end{table}

In our calculation of the $K^-pp$ system with both strong and Coulomb interactions taken into account, we find a deep-lying and short-lived state ($E_0^{s+C}$), widely known as a ``quasi-bound $K^- pp$ state caused by strong interactions'' and 
a high-lying and long-lived excited molecular state ($E_1^{s+C}$).

The primarily strong-induced quasi-bound $K^-pp$ state has an energy close to that of the ground state of the strong-only system. 
The energy difference between the two states is
$\Delta E_0^C = (1.69+i \,0.44)~\mathrm{MeV}$.
The $E_0^{s+C}$ state is the $E_0^s$ state perturbed by the Coulomb interaction, so that $\Delta E_0^C$ is the shift caused by the Coulomb interaction. The added interaction causes the state to become broader and more bound.

Similarly, the primarily Coulomb-induced molecular $K^-pp$ state has an energy close to that of the ground state of the Coulomb-only system, with an energy difference of
$\Delta E_{01}^{s}= (-0.55+i \,0.27)~\mathrm{keV}$. 
It is usually assumed that the $E_1^{s+C}$ state is the $E_0^C$ state perturbed by the strong interaction,
but similarly to the two-body $K^- p$ system it is rather the difference between the neighboring energy levels, since $E_0^C$ is the ground state of the $K^- pp$ system and $E_1^{s+C}$ is the excited state (see Ref.~\cite{kpPreprint}).

The behavior of the energies in the three-body $K^-pp$ system is similar to the behavior of the energies in the two-body $K^-p$ system.
It is seen from Table \ref{tbl:2b_results}, that
the energy difference between the two-body quasi-bound state and the strong-only state 
$\varepsilon_0^{s}  - \varepsilon_{1s}^{s+C}= (2.31+i0.90)~\mathrm{MeV}$ 
is larger in magnitude but directionally similar to the three-body shift $\Delta E_0^C$.
The difference between the atomic $K^- p$ state and the Coulomb $1s$ state
$ \varepsilon_{1s}^{C} - \varepsilon_{2s}^{s+C} = (-0.51 + 0.25i)~\mathrm{keV}$ 
is nearly identical to the energy difference between the corresponding three-body energy levels. This difference is usually considered to be the $1s$ level shift of kaonic hydrogen caused by the antikaon-nucleon strong interaction. 
As shown in Ref.~\cite{kpPreprint},
it is 
actually the difference between the neighboring energy levels of the $K^- p$ system since $\varepsilon_{1s}^C$ is the ground state and $\varepsilon_{2s}^{s+C}$ is the excited one.

We calculated the root-mean-square radius of the ground states of the strong-only, strong plus Coulomb, and Coulomb-only $K^-pp$ systems.
The radii of the strong-only and strong plus Coulomb  ground states are nearly identical, with $r^s_0= 0.92 \mathrm{~fm}$ and $r^{s+C}_0= 0.92 \mathrm{~fm}$, respectively.
The radius of the Coulomb-only state, $r^C_0= 174 \mathrm{~fm}$, is much larger.
If we assume the excited state of the strong plus Coulomb system is approximately the same size as the Coulomb-only ground state, $r^{s+C}_1 \approx r^C_0$,
then the primarily strong-induced quasi-bound $K^-pp$ state ($E_0^{s+C}$) is extremely compact, and the primarily Coulomb-induced molecular state ($E_1^{s+C}$) is far more spatially extended.

To validate the extrapolation scheme, we calculated  $E^{s+C}_0$ directly with $b^s$ and compared it to the value presented in Table \ref{tbl:3b_results}.
These results  differ by $2.42 - i8.33~\mathrm{keV}$.
This value corresponds to the difference $\Delta E_0^C|_{b=b^{s+C}} - \Delta E_0^C|_{b=b^{s}}$, and is negligible compared to the value $\Delta E_0^C = (1.69+i \,0.44)$ MeV, showing that the shift has nearly converged despite the fact that the energies have not converged.
We cannot make a similar comparison for $E^{s+C}_1$,
but the accuracy of Eq.~(\ref{eq:approx0}) and the stability of $\Delta E^s_{01}$ seen in Fig.~\ref{fig:shifts}
support the use of Eq.~(\ref{eq:approx1}) to obtain the energy and width of the Coulomb-induced molecular state. 

In Ref.~\cite{FineTune}, the binding energy and width of the $K^- pp$ ground state were calculated using the Faddeev-type AGS equations for the coupled $\bar{K}NN - \pi \Sigma N - \pi \Lambda N$ system without Coulomb interaction, for different models of the coupled-channel $\bar{K}N - \pi \Sigma - \pi \Lambda$ potential.
Here, we briefly discuss the differences between the present work and that work.
The most direct comparison is to the results obtained using the $V_{\bar{K}N-\pi \Sigma - \pi \Lambda}^{\rm 1,SIDD-A}$ coupled-channel potential with the same range parameter $\beta$ which gives the same energy for the $\Lambda(1405)$ resonance. 
The binding energy of the $K^- pp$ state in Table V of Ref.~\cite{FineTune} is $44.3$ MeV, which is larger than the absolute value of ${\rm Re }(E_0^{s})$ in Table \ref{tbl:3b_results}. 
In contrast, the corresponding width is $91.4$ MeV, which is smaller than twice the imaginary part of the $E_0^{s}$ energy. It was shown in Ref.~\cite{antiKNN_review} and references therein that the coupled-channel dynamics in the $\bar{K} NN$ system are very important. 
The primary differences arise from whether coupled-channel antikaon-nucleon potentials are or are not used in the three-body calculations. Additional differences come from adopting various nucleon-nucleon potentials, but it was shown in Ref.~\cite{PhysRevC.85.034001} that the influence of the $NN$ interaction model on the $\bar{K}NN$ results is quite small.

\section{Conclusion}
\label{conclusion}

We studied the $K^- pp$ molecular state caused primarily by the Coulomb interaction  together with the ``strong'' quasi-bound state of the same system.
We solved the Faddeev-Merkuriev integral equations in the Coulomb-Sturmian basis and located the three-body poles in the complex energy plane.
Both Coulomb and strong interactions were taken into account.
We modeled the $K^- p$ strong interaction with a complex one-term separable potential, while the $pp$ nuclear interaction was modeled with the MT-I Malfliet-Tjon potential.

The binding energy and width of the primarily Coulomb-induced molecular $K^- pp$ state are of the order of keV,
while the binding energy and width of the primarily strong-induced quasi-bound $K^- pp$ state are of the order of tens of MeV.
The energy shift between the strong-only ground state and the strong plus Coulomb ground state in the three-body $K^-pp$ system is similar to the corresponding shift in the two-body $K^-p$ system, with the two-body shift having a larger absolute value.
The energy difference between the Coulomb-only ground state and the strong plus Coulomb excited state in the three-body $K^-pp$ system is nearly identical to the corresponding difference in the two-body $K^-p$ system.
The quasi-bound $K^- pp$ state caused primarily by the strong interactions is an extremely compact ground state, while the molecular state is a far more spatially extended excited state.

\bibliographystyle{apsrev4-2}

\bibliography{kpp}

@article{bianchi1964convolution,
  title={A convolution integral for the resolvent of the sum of two commuting operators},
  author={Bianchi, L and Favella, L},
  journal={Il Nuovo Cimento (1955-1965)},
  volume={34},
  number={6},
  pages={1825--1828},
  year={1964},
  publisher={Springer}
}

@article{balian1969angular,
  title={Angular-momentum reduction of the Faddeev equations},
  author={Balian, R and Br{\'e}zin, E},
  journal={Il Nuovo Cimento B (1965-1970)},
  volume={61},
  number={2},
  pages={403--410},
  year={1969},
  publisher={Springer}
}

@book{faddeev2013quantum,
  title={Quantum scattering theory for several particle systems},
  author={Faddeev, L{\^u}dvig Dmitrievi{\v{c}} and Merkuriev, Stanislav Petrovich},
  volume={11},
  year={2013},
  publisher={Springer Science \& Business Media}
}

@article{merkuriev1980three,
  title={On the three-body Coulomb scattering problem},
  author={Merkuriev, S  {\relax P}},
  journal={Annals of Physics},
  volume={130},
  number={2},
  pages={395--426},
  year={1980},
  publisher={Elsevier}
}

@article{papp2002faddeev,
  title={Faddeev--Merkuriev equations for resonances in three-body Coulombic systems},
  author={Papp, Z and Darai, J and Nishimura, A and Hlousek, Z  {\relax T} and Hu, C-Y and Yakovlev, S  {\relax L}},
  journal={Physics Letters A},
  volume={304},
  number={1-2},
  pages={36--42},
  year={2002},
  publisher={Elsevier}
}

@article{papp2002resonant,
  title={Resonant-state solution of the Faddeev-Merkuriev integral equations for three-body systems with Coulomb potentials},
  author={Papp, Z and Darai, J and Hu, C-Y and Hlousek, Z  {\relax T} and K{\'o}nya, B and Yakovlev, S  {\relax L}},
  journal={Physical Review A},
  volume={65},
  number={3},
  pages={032725},
  year={2002},
  publisher={APS}
}

@article{PhysRevA.63.062721,
	author = {Papp, Z and Hu, C-Y and Hlousek, Z  {\relax T} and K\'onya, B and Yakovlev, S  {\relax L}},
	journal = {Phys. Rev. A},
	month = {May},
	number = {6},
	numpages = {11},
	pages = {062721},
	publisher = {American Physical Society},
	title = {Three-potential formalism for the three-body scattering problem with attractive Coulomb interactions},
	volume = {63},
	year = {2001}}

@article{PhysRevC.55.1080,
	author = {Papp, Z.},
	issue = {3},
	journal = {Phys. Rev. C},
	month = {Mar},
	pages = {1080--1087},
	publisher = {American Physical Society},
	title = {Three-potential formalism for the three-body Coulomb scattering problem},
	volume = {55},
	year = {1997}
}

@article{PhysRevC.54.50,
	author = {Papp, Z. and Plessas, W.},
	journal = {Phys. Rev. C},
	month = {Jul},
	number = {1},
	numpages = {6},
	pages = {50--56},
	publisher = {American Physical Society},
	title = {Coulomb-Sturmian separable expansion approach: Three-body Faddeev calculations for Coulomb-like interactions},
	volume = {54},
	year = {1996}}

@article{Malfliet1969,
  title   = {Solution of the Faddeev equations for the triton problem using local two-particle interactions},
  author  = {Malfliet, R. A. and Tjon, J. A.},
  journal = {Nuclear Physics A},
  volume  = {127},
  number  = {1},
  pages   = {161--168},
  year    = {1969},
  doi     = {10.1016/0375-4162(69)90775-1}
}

@article{Yamaguchi1954,
  title = {Two-Nucleon Problem When the Potential Is Nonlocal but Separable. I},
  author = {Yamaguchi, Yoshisuke},
  journal = {Physical Review},
  volume = {95},
  issue = {6},
  pages = {1628--1634},
  numpages = {0},
  year = {1954},
  month = {Sep},
  publisher = {American Physical Society},
  doi = {10.1103/PhysRev95.1628},
  url = {https://aps.org}
}

@article{PhysRevA.46.4437,
	author = {Papp, Zolt\'an},
	issue = {7},
	journal = {Phys. Rev. A},
	month = {Oct},
	pages = {4437--4439},
	publisher = {American Physical Society},
	title = {Use of Coulomb-Sturmian functions in calculating scattering quantities in Coulomb-like potentials},
	volume = {46},
	year = {1992}}

@article{PhysRevC.38.2457,
	author = {Papp, Zolt\'an},
	issue = {5},
	journal = {Phys. Rev. C},
	month = {Nov},
	pages = {2457--2460},
	publisher = {American Physical Society},
	title = {Potential separable expansion approach to scattering on Coulomb-like potentials},
	volume = {38},
	year = {1988}}

@article{PRADemir2006,
	author = {Demir, F. and Hlousek, Z  {\relax T} and Papp, Z.},
	issue = {1},
	journal = {Phys. Rev. A},
	month = {Jul},
	numpages = {4},
	pages = {014701},
	publisher = {American Physical Society},
	title = {Coulomb-Sturmian matrix elements of the Coulomb Green's operator},
	volume = {74},
	year = {2006}}

@book{papp2026mastering,
  title={Mastering Quantum Mechanics},
  author={Papp, Zolt{\'a}n},
  year={2026},
  publisher={Springer}
}

@article{Konya:1999by,
	archiveprefix = {arXiv},
	author = {K{\'o}nya, B. and L{\'e}vai, G. and Papp, Z.},
	eprint = {nucl-th/9908018},
	journal = {Phys.Rev.},
	pages = {034302},
	primaryclass = {nucl-th},
	slaccitation = {%%CITATION = NUCL-TH/9908018;%%},
	title = {{Continued fraction representation of Green's operators and unified description of bound, resonant and scattering states}},
	volume = {C61},
	year = {2000}}

@article{AY,
  title={(K-, $\pi$-) production of nuclear K̄ bound states in proton-rich systems via $\Lambda$∗ doorways},
  author={Yamazaki, Toshimitsu and Akaishi, Yoshinori},
  journal={Physics Letters B},
  volume={535},
  number={1-4},
  pages={70--76},
  year={2002},
  publisher={Elsevier}
}

@article{NGM,
  title={Faddeev calculation of a $K^-pp$ quasibound state},
  author={Shevchenko, N  {\relax V} and Gal, A and Mare{\v{s}}, J},
  journal={Physical Review letters},
  volume={98},
  number={8},
  pages={082301},
  year={2007},
  publisher={APS}
}

@article{NGMR,
  title={K{\={}} NN quasibound state and the K{\={}} N interaction: coupled-channels Faddeev calculations of the K{\={}} NN-$\pi$ $\Sigma$ N system},
  author={Shevchenko, N  {\relax V} and Gal, A and Mare{\v{s}}, J and Revai, J},
  journal={Physical Review C—Nuclear Physics},
  volume={76},
  number={4},
  pages={044004},
  year={2007},
  publisher={APS}
}

@article{IS,
  title={Strange dibaryon resonance in the K{\={}} NN-$\pi$ YN system},
  author={Ikeda, Y and Sato, T},
  journal={Physical Review C—Nuclear Physics},
  volume={76},
  number={3},
  pages={035203},
  year={2007},
  publisher={APS}
}

@article{FineTune,
  title={Fine-tuning of the K{\={}} NN and K{\={}} K{\={}} N quasibound state calculations},
  author={Shevchenko, N {\relax V}},
  journal={Physical Review C},
  volume={112},
  number={6},
  pages={064007},
  year={2025},
  publisher={APS}
}

@article{E15_JPARC1,
  title={“K- pp”, a K‾-meson nuclear bound state, observed in 3He (K-, $\Lambda$p) n reactions},
  author={Ajimura, S and Asano, H and Beer, G and Berucci, C and Bhang, H and Bragadireanu, M and Buehler, P and Busso, L and Cargnelli, M and Choi, S and others},
  journal={Physics Letters B},
  volume={789},
  pages={620--625},
  year={2019},
  publisher={Elsevier}
}

@article{SIDDHARTA,
  title={A new measurement of kaonic hydrogen X-rays},
  author={Bazzi, M and Beer, G and Bombelli, Luca and Bragadireanu, AM and Cargnelli, M and Corradi, G and Curceanu, C and Fiorini, C and Frizzi, T and Ghio, F and others},
  journal={Physics Letters B},
  volume={704},
  number={3},
  pages={113--117},
  year={2011},
  publisher={Elsevier}
}

@article{kpPreprint,
  title={Level rearrangement in K-p system},
  author={Massimino, T and Shevchenko, N  {\relax V} and Papp, Z and Revai, J},
  journal={arXiv preprint arXiv:2606.21261},
  year={2026}
}

@ARTICLE{1977JMP....18.2316H,
	author = {{Hill}, Robert Nyden},
	title = "{Proof that the H$^{-}$ ion has only one bound state. Details and extension to finite nuclear mass}",
	journal = {Journal of Mathematical Physics},
	year = 1977,
	month = dec,
	volume = {18},
	number = {12},
	pages = {2316-2330},
	doi = {10.1063/1.523241},
	adsurl = {https://ui.adsabs.harvard.edu/abs/1977JMP....18.2316H}
}

@article{Richard:2016eca,
	author = "Richard, Jean-Marc and Fayard, Claude",
	title = "{Level rearrangement in exotic-atom-like three-body systems}",
	eprint = "1609.09350",
	archivePrefix = "arXiv",
	primaryClass = "cond-mat.quant-gas",
	doi = "10.1016/j.physleta.2017.08.021",
	journal = "Phys. Lett. A",
	volume = "381",
	pages = "3217--3221",
	year = "2017"
}

@article{Moriya:2019ans,
    author = "Moriya, H. and Horiuchi, W. and Richard, J. -M.",
    title = "{Three-body correlations in mesonic-atom-like systems}",
    eprint = "1911.05972",
    archivePrefix = "arXiv",
    primaryClass = "nucl-th",
    doi = "10.21468/SciPostPhysProc.3.051",
    journal = "SciPost Phys. Proc.",
    volume = "3",
    pages = "051",
    year = "2020"
}

@article{antiKNN_review,
  title = {Three-Body Antikaon–Nucleon Systems},
  author = {Shevchenko, N. V.},
  journal = {Few-Body Systems},
  volume = {58},
  pages = {6},
  year = {2017},
  doi = {10.1007/s00601-016-1170-5},
  url = {https://doi.org/10.1007/s00601-016-1170-5}
}

@article{PhysRevC.85.034001,
  title = {One- versus two-pole $\overline{K}N$-$\ensuremath{\pi}\ensuremath{\Sigma}$ potential: ${K}^{\ensuremath{-}}d$ scattering length},
  author = {Shevchenko, N. V.},
  journal = {Phys. Rev. C},
  volume = {85},
  issue = {3},
  pages = {034001},
  numpages = {16},
  year = {2012},
  month = {Mar},
  publisher = {American Physical Society},
  doi = {10.1103/PhysRevC.85.034001},
  url = {https://link.aps.org/doi/10.1103/PhysRevC.85.034001}
}

\end{document}